\documentclass[twocolumn]{aa}
\usepackage{graphicx}
\begin{document}
\title{High resolution spectroscopy of Brown Dwarfs in Taurus}

\author{David Barrado y Navascu\'es.\altaffilmark{1}}

   \author{David Barrado y Navascu\'es.
          }

   \offprints{D. Barrado y Navascu\'es}

   \institute{Laboratorio de Astrof\'{\i}sica Espacial y F\'{\i}sica Fundamental,
INTA, P.O. Box 50727, E-2808 Madrid, SPAIN 
              \email{barrado@laeff.esa.es}
             }

   \date{Received ...; accepted ...}

   \abstract{
We present high resolution optical spectroscopy of three
candidate members of the Taurus-Auriga star forming region. Based 
on the spectral type, the strength, profile and width of the H$\alpha$ line,
 the detected lithium at 6708 \AA, the location of these objects in a H-R
diagram and the comparison with similar objects belonging to young
stellar associations, we determine that they are bona fide members of the SFR, 
with about $\sim$3 Myr, have masses at or below the substellar limit and, at least in 
one case, there is active accretion from a circum(sub)stellar disk.
This result suggests that high mass brown dwarfs go through a Classical TTauri
phase and form like stars, from colapse and fragmentation of a molecular cloud.
   }

   \maketitle
%

\section{Introduction}

Brown dwarfs, objects unable to fuse hydrogen in an stable manner 
(i.e., with masses below at about 0.072 M$_\odot$, Baraffe et al. 1998),
 pose an important problem to the theory 
of stellar formation. Several formation mechanisms have been proposed, 
including formation like a star (from collapse and fragmentation of a 
molecular cloud, Padoan P., \& Nordlund 2004) to a planet-like process 
(from a circumstellar disk) or as stellar ``embryos'', ejected from multiple systems
before they are able to accrete enough matter (Reipurth \& Clarke 2001;
Bate et al. 2002). 
These proposed mechanisms have different implication in their
 formation, their evolution and their properties.
In particular, if brown dwarfs are  created like stars, 
it seems that they should go through an phase of active accretion  from
a circum(sub)stellar disk, such as  low mass stars do
 (Shu et al$.$ 1987).  Actually, since the 
last couple of years, different groups have presented indirect and
direct evidences of active accretion in a handful of young brown dwarfs 
belonging to several nearby star forming regions (SFR), open clusters and 
moving groups 
(Fern\'andez \& Comer\'on 2001;
Muench et al. 2001; 
Natta \& Testi 2001;
Natta et al. 2001, 2002;
Testi et al. 2002;
Jayawardhana et al. 2002ab, 2003ab;
Muzerolle et al. 2003;
Barrado y Navascu\'es et al. 2002, 2003, 2004;
Barrado y Navascu\'es \& Mart\'{\i}n 2003;
Mohanty \& Basri 2003; Mohanty, Jayawardhana \& Barrado y Navascu\'es 2003;
Comer\'on et al. 2003)

In this paper, we enlarge the sample by collecting high resolution spectroscopy of
three proposed members of the Taurus-Auriga SFR, located at 140 pc and about 1-3 Myr
(for an update on Taurus, see Luhman et al. 2003 and references therein).

\section{Observations}

Our Taurus targets were selected from Brice\~no et al. (2002).
In that paper, a new sample of nine mid- to late-M candidate members 
were presented, based on I$c$ and $z$' photometry (plus 2MASS near IR data) and
low resolution spectroscopy. 
We have collected  high resolution spectra of a third of them
--the high mass brown dwarfs,  whose spectral type is between M5.75 and M7.5--
with the Magellan I 6.5m telescope and the MIKE echelle spectrograph
on 2002, Dec 11-14th. Although three objects are a small number,
they represent a significant fraction of the high mass brown dwarfs
belonging to this association discovered so far (about 12 in the quoted spectral range).
 Therefore, they can provide some hints about their formation mechanism. 

Additional details about the observations can be found in 
Barrado y Navascu\'es, Stauffer \& Jayawardhana (2004), where we analyze a 
sample of very low mass stars and brown dwarfs belonging to the $\sim$5 Myr 
cluster associated to the $\lambda$ Orionis star.
In order to optimize the signal-to-noise, we binned the
 data during the read-out to two by two pixels in the spatial and the
spectral direction, respectively, yielding a spectral resolution of
R=25,000 ($\sim$0.25 \AA), with a 0.75 arcsec slit. 
The spectral range  of our spectra is 4500--7250 \AA.

\section{Analysis and Discussion}

\subsection{Spectral types}

Spectral types were derived by comparing with several spectral
templates, by  using the order around 7050 \AA,
corresponding to a TiO band,
following Mohanty et al. (2004), since this range is very sensitive to 
effective temperature. Errors can be estimated as half a subclass.
Our values are very close to those obtained by 
Brice\~no et al. (2002) from low resolution spectra (see Table 1).

\setcounter{figure}{0}
    \begin{figure}
    \centering
    \includegraphics[width=8.2cm]{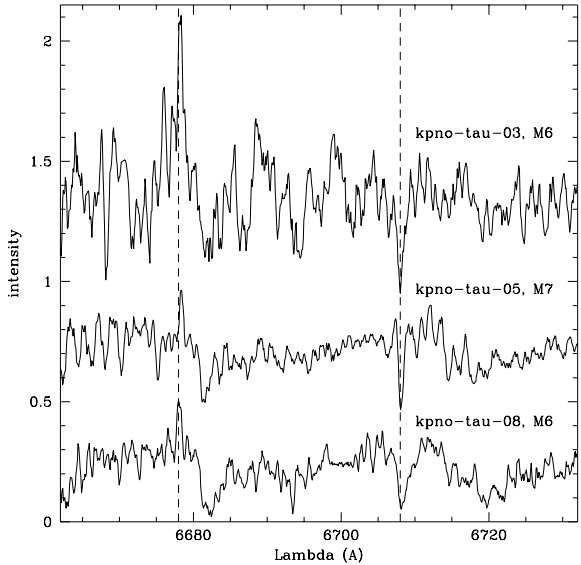}
 \caption{
 Spectra around HeI6678 \AA{} and LiI6707.8 \AA.
  }
 \end{figure}

\subsection{On the lithium abundance, age, and mass}

We have detected lithium 6707.8 \AA{} in two out of the three
targets
(KPNO-Tau-05 and KPNO-Tau-08, Figure 1).
The spectrum of KPNO-Tau-03 has worse quality.
With some caveats, the visual inspection indicates
 that this feature is also present.
Note that   Brice\~no et al. (2002)  states that they detected
lithium  with their low resolution spectrum (at higher signal-to-noise ratio).
This element is easily destroyed in the stellar interior, being its surface abundance 
dependent on mass and age (as well as other second order parameters).
In fact, brown dwarfs which are more massive than about 0.060 M$_\odot$ do deplete lithium
in a time scale of few tens of million years (D'Antona \& Mazzitelli 1994; 
Burrows et al. 1997; Baraffe et al. 1998). The Taurus-Auriga complex has an age between
1  and 3 Myr. Since our three candidate members have spectral types
between M6 and M7, they should have masses equal or larger than 0.06 M$_\odot$, according to 
Baraffe et al. (1998). Therefore, the detection of this alkali clearly indicates
that these objects are in the  pre-Main Sequence (PMS). Moreover, from the statistical 
point of view, the likelihood of having three late-M, PMS interloper whose spectral and 
photometric properties coincide with the Taurus sequence is negligible. Therefore, we have to
conclude that they, indeed, belong to the association.

\setcounter{figure}{1}
    \begin{figure}
    \centering
    \includegraphics[width=8.2cm]{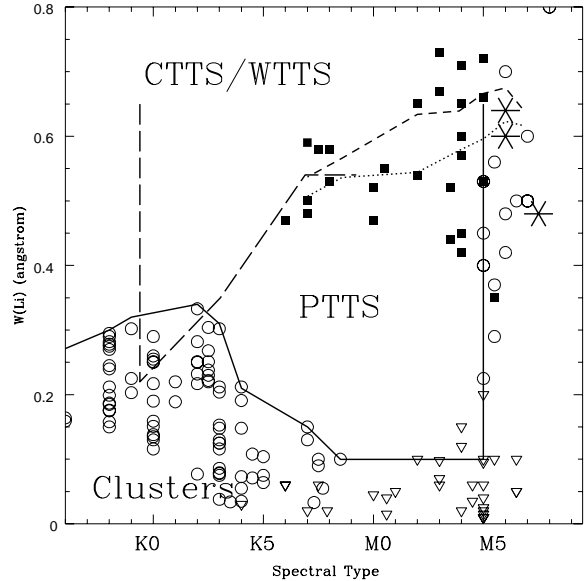}
 \caption{
Lithium equivalent width versus the spectral type.
The solid line corresponds to the upper envelope of the values
measured in young open clusters.
The long-dashed line delimits the areas for weak-line and post-T~Tauri
stars (adapted from Mart\'{\i}n 1997 and Mart\'{\i}n \& Magazz\`u 1999).
Short-dashed and dotted lines correspond to the
cosmic abundances --A(Li)=3.1-- from gravities of Logg=4.5 and 4.0,
respectively 
(curves of growth from Zapatero Osorio et al$.$ 2002).
All Pleiades and IC2391  members with measured lithium equivalent width are shown
as open circles and triangles --upper limits--.
 Note the lithium depletion boundary at  $\sim$M5.5.
 Sigma Orionis low mass stars and brown dwarfs appear
as solid squares (Zapatero Osorio et al$.$ 2002). 
The big asterisks represent KPNO-Tau-3, 5 and 8. 
  }
 \end{figure}

Figure 2 displays values of the lithium equivalent width (W) measured for 
our targets (large asterisks), pre-main sequence members of the $\sim$5 Myr
Sigma Orionis cluster (Zapatero Osorio et al. 2002, solid squares),
and stellar and substellar members of IC2391 --53 Myr-- and the Pleiades --125 Myr--
from Barrado y Navascu\'es et al. (1999, 2004), Soderblom et al. (1993),
Garc\'{\i}a-L\'opez et al. (1994), Jones et al. 1996, Stauffer et al. (1998) and
Jeffries et al. (1999).
The solid line describe the maximum W(Li) measured in cluster stars.
The long-dashed curve is an update version of the criterion used by Mart\'{\i}n (1997)
and Mart\'{\i}n \& Magazz\`u (1999) to distinguish between Weak-line and post- TTauri stars.
Finally, the dotted and short dashed lines correspond to the Zapatero Osorio et al. (2002)
curves of growth for Log\,g=4.0 and 4.5 and an abundance of A(Li)=3.1
(i.e., cosmic abundance). We have labeled in the plot different sections.
The W(Li) for KPNO-Tau-5 is lower than the measured values in the other two Taurus objects.
This fact might be due to the presence of
optical veiling, about r(6700)$\sim$0.2, although we do not think this is the case. 
On one hand, no other sign of accretion has been found in this object
(see next section).
Moreover, this equivalent width  is 
compatible with the dispersion found in Sigma Orionis
and the  --undepleted- values characteristic of cluster brown dwarfs.
For the other two objects, no veiling seems to be present either.
In the lithium equivalent width dispersion is real both in Taurus and Sigma Orionis
cluster
(the same effect is present in the Lambda Orionis association, 
Barrado y Navascu\'es, Stauffer \& Jayawardhana 2004), it might 
imply diferences in the surface abundances due to additional mixing mechanisms
in the stellar and substellar interior, different to pure convection
 (such as those proposed in solar mass stars),
differences in the structure due to different stellar and substellar 
parameters (such as fast rotation in the case of pre-Main sequence stars, 
Mart\'{\i}n \& Claret  1996),
or to induced effects related to  activity and/or rotation over the spectral feature
itself or the surrounding continuum (see the discusion in Barrado y Navascu\'es et al. 2001 
and references therein).

 As a summary, based on these data, mainly from the detection  of lithium
and accretion when present (see next section), 
we can conclude that these three objects are young and, indeed,
belong to the Taurus region.
In the case of KPNO-Tau-5, due to its spectral type (M7) its 
substellar nature is well established. The other two are located at the substellar
borderline and their nature is not so firmly established due to uncertainties in 
the models and the spectral type determination.  

Once membership to the stellar association has been proved, we have 
derived the bolometric 
magnitudes from $Ic$ and $Ks$, the distance modulus 
$(m-M)_0$=5.731 
(140 pc, Kenyon, Dobrzycka \& Hartmann 1994), 
the reddening derived by Brice\~no et al. (2002)
and the bolometric corrections by Comer\'on et al. (2000) 
and Tinney et al. (1993)
 for these two bands ($Ic$ and $Ks$, respectively).  
Masses were computed based on
models by Baraffe et al. (1998). 
Effective temperatures were obtained
using several scales, namely
Luhman (1999) for intermediate gravity
 and Leggett (2000, 2001).
 All the measured and derived 
 values are listed in Table 1.

\setcounter{figure}{2}
    \begin{figure}
    \centering
    \includegraphics[width=8.2cm]{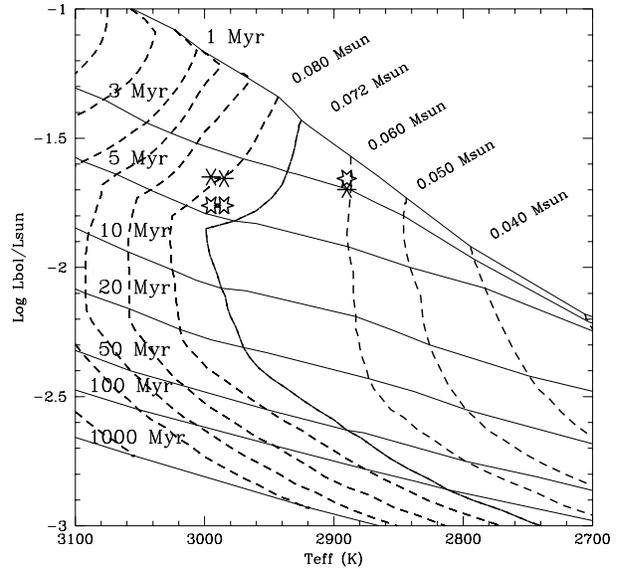}
 \caption{HR diagram of KPNO-Tau-3 and 8 (on top of each other)
and KPNO-Tau-5. The values represented by asterisks and open stars 
were derived from $Ic$ and $Ks$ magnitudes, respectively. }
 \end{figure}

Figure 3 displays a HR diagram. 
The isochrones and evolutionary tracks 
--solid and dashed lines, respectively-- are
from Baraffe et al. (1998).
Asterisks and  open stars correspond to bolometric luminosities 
obtained from $Ic$ and $Ks$, respectively.
For this particular diagram, we made use of the
effective temperature  scale by Luhman (1999). 
Other temperature scales, such as that from
 Leggett et al. (2001),
would shift the location of these three objects (two of them are
almost on top of each other, the small shift in Teff is arbitrary)
 to the right hand-side, making them younger
and less massive. Note, however, that Luhman' scale has been tunned specifically
for the model we are using here.
In any case, regardless the election of models (alternative models are, for instance, 
Burrows et al. 1997; D'Antona \& Mazzitelli 1994, 1997, 1998;
Chabrier  et al. 2000; Baraffe et al. 2002),
bolometric corrections and effective temperature scale, 
these members are at or below the 
substellar frontier and have an age of about 3 Myr.

\setcounter{figure}{3}
    \begin{figure}
    \centering
    \includegraphics[width=8.2cm]{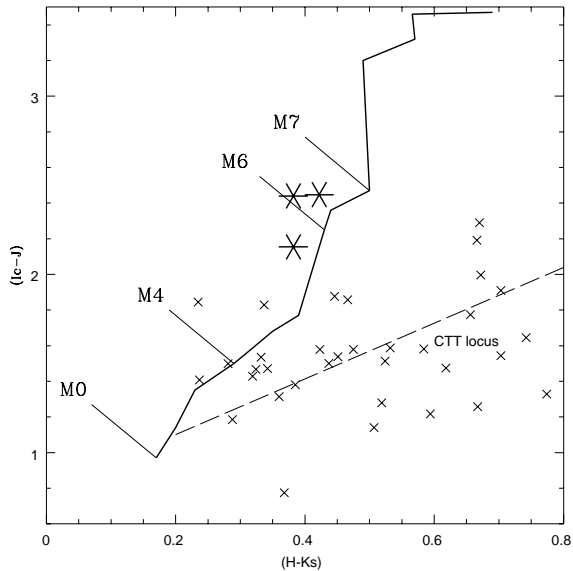}
 \caption{Color-color diagrams.
The position of the Taurus BD candidates are indicated with large 
asterisks.
 Crosses indicate the position of classical TTauri stars belonging to Orion
stellar population (Herbig \& Bell 1988).
The thick-solid and dashed lines correspond to the locii of the main
 sequence stars (from Bessell \& Brett 1988; Kirkpatrick et al$.$
 2000; Leggett et al. 2001) 
and CTT stars (Meyer et al$.$ 1997, Barrado y Navascu\'es et al. 2003), respectively.   }
 \end{figure}

\subsection{Near infrared photometry, H$\alpha$ emission and accretion}

None of our three targets   seems to have neither the presence of
forbidden lines, characteristics of outflows, nor  near infrared
infrared excesses, coming from an accretion disk.
Figure 4 compares the colors $(Ic-J)$ and $(H-Ks)$,
 and includes  the loci for dwarfs --Bessell \& Brett 1988; 
Kirkpatrick et al. 2000; Leggett et al. 2001)-- and Classical TTauri stars
(Meyer et al. 1997, Barrado y Navascu\'es et al. 2003)
and Classical TTauri stars from Orion (Herbig \& Bell 1988).
As can be seen, the Taurus objects have photometric properties
--data from 2MASS-- similar to those MS stars of similar spectral type
and, as stated in the previous paragraph, no near IR excess is seen.
This fact, by itself, cannot  prove or disprove the presence of a
 circum(sub)stellar disk, since a hole can be present  or the disk 
temperature can be very cold. Moreover, the IR excess might depend on the orientation
of the disk. See, for example, the case of LS-RCrA~1, M6.5 brown dwarf belonging
 to RCrA dark cloud (Barrado y Navascu\'es, Mohanty \& Jayawardhana 2004).
Additional observations and longer wavelengths --more sensitive to cooler disks-
 both from ground-based telescopes
and space-borne instruments such as those in Spitzer Space Telescope, 
can help to shed some light in this issue.

\setcounter{figure}{4}
    \begin{figure}
    \centering
    \includegraphics[width=8.2cm]{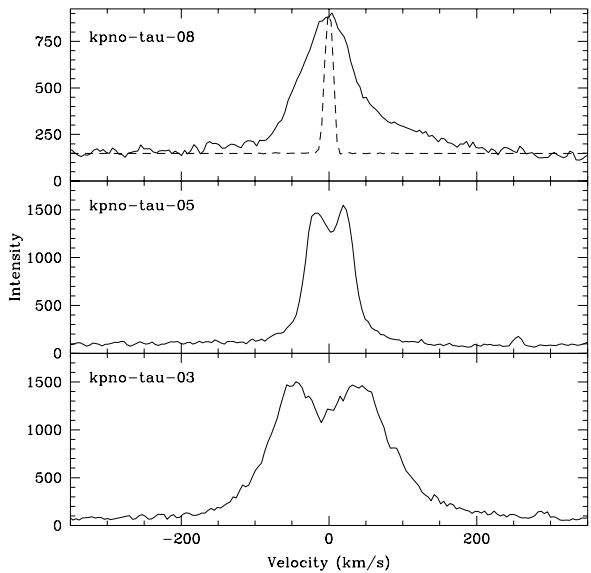}
 \caption{
 H$\alpha$ profiles. The instrumental profile is included 
as a dashed line (top spectrum).
  }
 \end{figure}

We have detected and measured H$\alpha$ in emission for these three objects.
The profiles of this feature are displayed in Figure 5.
 Note the possible asymmetry 
in KPNO-Tau-8, the double peak in KPNO-Tau-5 and KPNO-Tau-3,
 and the width of the line
(310 km/s), typical of accreting brown dwarfs
(White \& Basri 2003; Jayawardhana et al. 2003). 
In fact, this last object is above the criterion defined 
by Barrado y Navascu\'es \& Mart\'{\i}n (2003) which discriminate between 
accreting and non accreting objects, as Figure 6 clearly indicates.
We note, however, that this criterion depends on low resolution 
spectroscopy, which normally yields larger equivalent widths compared with 
higher resolution data.
In this diagram, we have included Classical and Weak-line TTauri
stars belonging to Taurus as solid and open circles. 
Members without classification are included as crosses.
Our three targets appear as large asterisks. Overlapping squares and big circles
denote those members with forbidden emission lines and near-infrared excesses, 
respectively.
This large W(H$\alpha$) agree with the fact that we also detect
 HeI6678 \AA, another accretor indicator (see Figure 1).

\setcounter{figure}{5}
    \begin{figure}
    \centering
    \includegraphics[width=8.2cm]{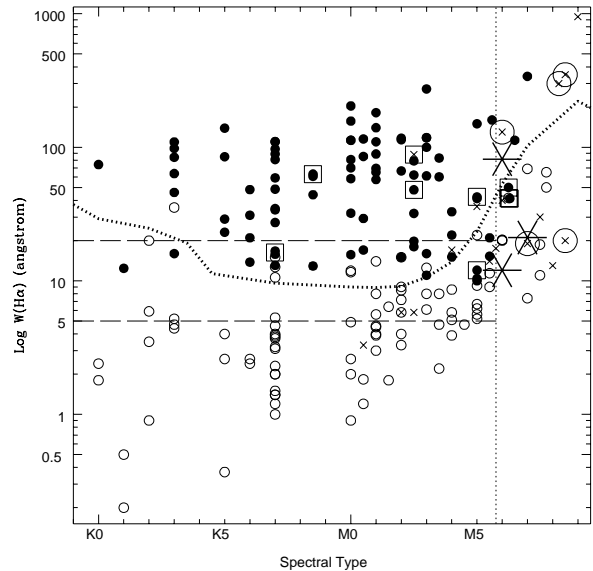}
 \caption{H$\alpha$ equivalent widths for members of the Taurus SFR.
Solid  and open circles correspond to Classical and TTauri stars.   
Objects with no classification are shown as crosses. 
Large open circles or square  represent objects with mid-IR excesses
and forbidden lines, respectively.
The three objects studied here are displayed as large asterisks.  
The dotted, bold curve is the saturation  criterion, 
whereas two previously proposed criteria (5 and 20 \AA) to separate  
CTTS and WTTS  
are included as long-dashed, thin horizontal segments. 
The vertical dotted segment denotes the location of the substellar  
frontier.  
}
 \end{figure}

Active accretion in at least one Taurus member whose mass is close or below the
substellar limit is important for several reasons. First, although the sample
studied in this paper is very small, it suggests that a significant fraction
of the very low mass stars and high mass brown dwarfs might harbor an accretion disk.
Second, the new data amass additional evidence for accretion in the substellar
domain. To the best of our knowledge, there are only three papers dealing with 
high resolution spectra in Taurus brown dwarfs  
(White \& Basri 2003; Muzerolle et al. 2003;  and Jayawardhana  et al. 2003).
The first work found three accretors in  a sample of ten very low mass stars and 
brown dwarfs, although the less massive, a M6.5 (GM~Tau), 
  has a mass just above the substellar limit.
 The second paper includes four objects  whose spectral type is M6 or M7.
None of them seem to undergo active accretion based on the width of H$\alpha$, although 
the authors claim that one of them, namely MHO-5, is accreting based on the 
detection of forbidden lines of oxygen at 6300 and 6363 \AA{ } and CaII IRT.
Moreover, three out of the four were observed at moderate resolution, including MHO-5
 (R$\sim$8000).
Regarding the later study, the four M7-M7.75 brown dwarfs discussed there, with masses
down to 0.05 M$_\odot$ (again, using Baraffe et al. 1998 models), 
do not show accretion either.
Muzerolle et al. (2003) also analyzed 
 seven Taurus members whose spectral types are M4.75--M5.75 
(just above or at the substellar borderline).
 Three out of these
seven are accreting, based on the width of H$\alpha$ and other spectral indicators.
Therefore, up to date, our study  presents the 
only  brown dwarf belonging to Taurus (kpno-tau-03), at the substellar limit, 
which has been proved to be accreting. 

In conjunction with the studies quoted in the previous  paragraph,
 our results indicate that 
about 10\% of the Taurus brown dwarfs (one out of 11) is actively accreting.
 This fraction is much smaller than
the accretion occurrence in low mass stars in the association or
 the estimate for the substellar
 domain (about 50\%) based purely on the strength of H$\alpha$
 measured in low resolution 
spectra (Barrado y Navascu\'es \& Mart\'{\i}n 2003).
The statistical criterion defined in this last work is based on the saturation
of the activity, as measured in several young open clusters
(namely IC2391, Alpha Per and the Pleiades, with ages ranging from about 50
to 125 Myr). The discrepancy in the fraction of accreting brown dwarfs  might imply
that there is an additional source of flux in H$\alpha$ line  or that the 
accreting criterion based on the H$\alpha$ width at 10\% of the maximum intensity,
 as defined by Jayawardhana et al. (2003), i.e. 200 km/s, is too restrictive.

The detection of accretion in substellar objects indicates that they 
undergo a phase 
similar to Classical TTauri stars (for a review, see Bertout 1989 
or Appenzeller \& Mundt 1989).
 A handful of other brown dwarfs belonging to
other young stellar association, with ages  ranging from 1 to 10 Myr, have been 
observed with high resolution spectroscopy and the presence of accretion
 confirmed. In at least one, LS-RCrA~1 
(Fern\'andez \& Comer\'on 2001; Barrado y Navascu\'es, Mohanty \& Jayawardhana 2004),
outflows have been seen, by means of the detection of intense, narrow forbidden lines.
In a previous paper we have named them Classical TTauri substellar analogs (CTTSA).
Therefore, we can conclude, at least in the case of
high mass brown dwarfs (Mas$\sim$0.072--0.04 M$_\odot$), that they are formed as
low mass stars, by fragmentation and collapse of the original molecular cloud.
Of course, additional studies are needed to confirm this preliminary conclusion, 
in particular high resolution imaging in the near and mid- infrared, as well as in the optical.
Narrow band imaging might show whether these objects present outflows similar to those
observed in some Classical TTauri stars.

\section{Summary}

We have collected high resolution spectroscopy for the members of the 
Taurus-Auriga SFR. They have  M6-M7 spectral types with 
 moderate to intense H$\alpha$ emission, with profiles that, at least in
one case, can be classified as typical of an accretor.
We detect lithium too, which indicate that they are in the pre-main sequence
(more strictly, the equivalent width  imposes a maximum age of about 30 Myr, 
based on Baraffe et al. 1998). 
Moreover, at least one, possibly two of them, are accreting material, as shown 
by the H$\alpha$ profile, providing an additional, stronger
constrain to the age, about 10 Myr.
Therefore, a minimum of a10\% (one out of eleven) of the high mass brown dwarfs 
known in the cluster show accretion (i.e., circumsubstelalr disks).
Note, however, that these objects do not have near-infrared excess 
and do not show forbidden lines, although HeI6678 \AA, another signpost
of accretion, appears in the spectrum of one of them, possibly in the other two.
Since the likelihood  of having a stellar 
association younger than 30 Myr in the same line-of-sight as
Taurus, we can safely conclude that these objects belong to this 
SFR. 
Based on the new spectral types and magnitudes in the optical and infrared,
we derived masses around the substellar limits and an age close to 3 Myr.
Finally, the detection of accretion in the substellar domain provides another
clue about the formation mechanism of high mass brown dwarfs, suggesting that
they form as stars, by fragmentation and collapse of a molecular cloud.

\begin{table*}
\caption[]{Data for the three Taurus low mass members.}
\begin{tabular}{cccc}
\hline
                            &\multicolumn{3}{c}{KPNO-Tau}\\
\cline{2-4}
                            &   \#3            &    \#5   &  \#8 \\
\hline                                                                                                    
$Ic$                        & 15.77            &15.08            &15.10            \\
$J$                         & 13.323$\pm$0.022 &12.640$\pm$0.024 &12.946$\pm$0.022 \\
$H$                         & 12.501$\pm$0.022 &11.918$\pm$0.024 &12.367$\pm$0.021 \\
$Ks$                        & 12.079$\pm$0.021 &11.536$\pm$0.018 &11.985$\pm$0.022 \\
Sp.Type                     & M6 (M6)	       & M7 (M7.5)       & M6 (M5.75) \\
Width H$\alpha$ (km/s)      & 310    	       & 125             & 240$^1$\\
W(H$\alpha$) (\AA)          &81.4$\pm$5.7 (130)&21.1$\pm$0.9 (30)& 12.0$\pm$1.0 (17.5)\\
W(HeI6678)   (\AA)          & 3.5: (6.5)       &  0.2:           & 0.2: \\
W(Li)        (\AA)          & 0.60 (Yes)       &  0.48           & 0.78 \\
Log Lum(bol)/L$_\odot$      & -1.66/-1.76$^2$  & -1.70/-1.66$^2$ & -1.65/-1.76$^2$    \\
Mass (M$_\odot$)            & 0.07$^3$         & 0.06            & 0.07$^3$     \\
Log {L(H$\alpha$)}/{L(bol)} & -2.87            &  -3.70          & -3.70  \\
Teff (Luhman 1999)          & 2990             & 2890            & 2990 \\       
\hline
\end{tabular}
$\,$  \\
Values in parenthesis --from low resolution spectroscopy--
  from Brice\~no et al. (2002). A double colon implies a large
uncertainty in the measurement.\\
$^1$ 180 km/s without the asymmetry. \\
$^2$ From $Ic$ and $Ks$, respectively. \\
$^3$ 0.08 M$_\odot$ from the HR diagram. 
The listed values are from $Ic$ and Baraffe et al. (1998)\\
\end{table*}

\begin{acknowledgements}
  This work has been supported by the Spanish ``Programa Ram\'on y Cajal'' and the
PNAyA AYA2003-05355. Comments by M.R. Zapatero Osorio are appreciated.
\end{acknowledgements}


\end{document}